\documentclass[twocolumn,aps,prl,showpacs,amsmath,superscriptaddress,longbibliography,notitlepage]{revtex4-2}
\usepackage{amssymb}
\usepackage{mathrsfs}
\usepackage{graphicx}
\usepackage{float}
\usepackage[caption=false]{subfig}

\usepackage[normalem]{ulem}
\usepackage{verbatim}
\usepackage{xcolor}
\usepackage{bm}
\usepackage{makecell}

\begin{document}

\title{Saturation Dynamics in Non-Hermitian Topological Sensing Systems}
\author{S. M. Rafi-Ul-Islam}
\email{rafiul.islam@u.nus.edu}
\affiliation{Department of Electrical and Computer Engineering, National University of Singapore, Singapore 117583, Republic of Singapore}
\author{Zhuo Bin Siu}
\email{elesiuz@nus.edu.sg}
\affiliation{Department of Electrical and Computer Engineering, National University of Singapore, Singapore 117583, Republic of Singapore}
\author{Md. Saddam Hossain Razo}
\email{shrazo@u.nus.edu}
\affiliation{Department of Electrical and Computer Engineering, National University of Singapore, Singapore 117583, Republic of Singapore}
\author{Mansoor B.A. Jalil}
\email{elembaj@nus.edu.sg}
\affiliation{Department of Electrical and Computer Engineering, National University of Singapore, Singapore 117583, Republic of Singapore}

\begin{abstract}
A class of non-Hermitian topological sensors (NTOSs) was recently proposed in which the NTOS comprises a non-Hermitian Su-Schrieffer-Heeger chain with a measurant-dependent coupling between the two ends of the chain. The smallest eigenenergy of the system, which serves as the readout signal, has an exponential dependence on the system size at small system sizes but saturates above a critical size. In this study, we further elucidate the dependence of the sensor sensitivity and saturation behavior on the system parameters. We explain how the behavior of the NTOS is characterized by a winding number, which indicates whether the smallest eigenenergy decreases to zero exponentially with the system size or grows exponentially up to a critical size. Interestingly, we further show that by imposing unidirectionality on the coupling between the two ends of a sensor, we can flip the size dependence of the smallest eigenenergy value from an exponentially increasing trend to an exponentially decreasing one. Our findings provide important insights into the saturation phenomenon and the impact of terminal couplings on the sensing characteristics of NTOSs.
\end{abstract}
\maketitle

\section{Introduction}
The field of topological physics has witnessed remarkable advancements in recent years \cite{lu2016topological,rafi2020realization,ma2019topological,rafi2024chiral,rafi2020strain,roy2009topological,rafi2022valley,gong2018topological}, particularly in the field of non-Hermitian systems \cite{esaki2011edge,rafi2021topological,liu2019second,rafi2021non,lieu2018topological,yin2018geometrical,yuce2015topological}. Non-Hermitian topological systems exhibit intriguing properties \cite{li2020critical,rafi2022critical,okuma2020topological,rafi2022interfacial,de2022non,budich2020non,rafi2022critical,mcdonald2020exponentially} such as non-trivial boundary states \cite{zou2021observation,fan2022hermitian,koch2020bulk}, non-Hermitian skin effects \cite{okuma2020topological,zhang2022review,kawabata2020higher,zhang2022universal}, and novel topological invariants \cite{yao2018edge,ghatak2019new,song2019non}. These systems have garnered significant attention owing to their potential applications across a wide range of fields including photonics \cite{feng2017non,parto2020non,wang2021topological}, topolectrical (TE) circuits \cite{lee2018topolectrical,rafi2020anti,hofmann2020reciprocal,rafi2022unconventional,imhof2018topolectrical,rafi2023valley,kotwal2021active,rafi2022type,sahin2023impedance},  condensed matter physics \cite{martinez2018topological,pan2020non,okuma2021non}, and quantum computing \cite{bender2007faster,zhao2021anomalous,gopalakrishnan2021entanglement}.

One promising application of non-Hermitian topological systems is  the development of highly sensitive sensors \cite{budich2020non,lau2018fundamental,mcdonald2020exponentially,hokmabadi2019non,koch2022quantum,yuan2023non}. These sensors exploit the unique properties of non-Hermitian systems states to achieve enhanced sensing capabilities, which surpass the limits of conventional sensing technologies \cite{budich2020non,guo2021exact,lau2018fundamental,mcdonald2020exponentially}. In particular, the exponential amplification of the sensing outputs, known as the ``exponential sensing effect", has attracted considerable interest \cite{budich2020non,lau2018fundamental}. This effect allows for exponentially enhanced signal detection, which can potentially lead to the detection of weak signals with high precision.

\begin{figure}[htp]
\centering
\includegraphics[width=0.48\textwidth]{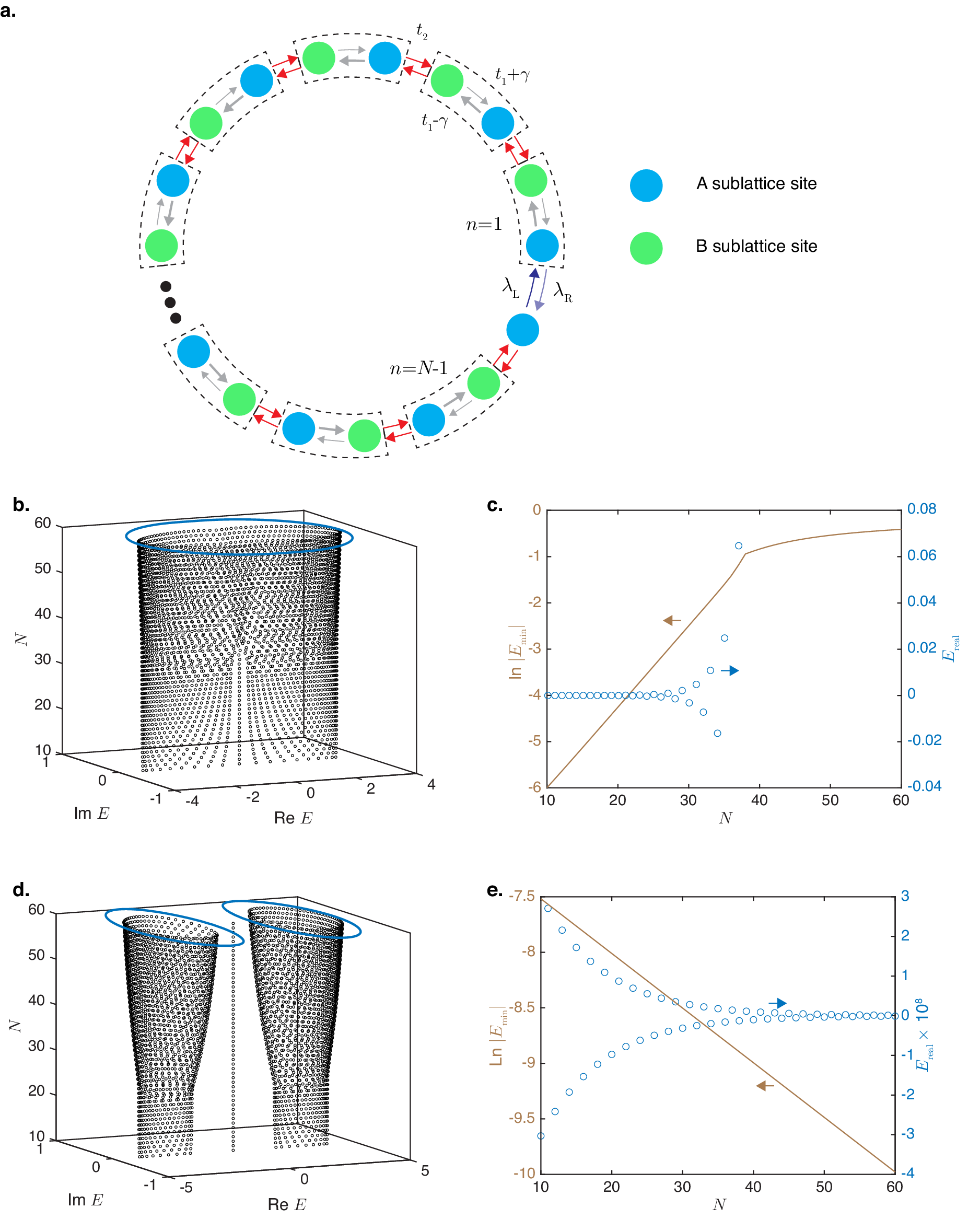}
\caption{\textbf{a}. A schematic representation of the non-Hermitian SSH chain considered comprising $N-1$ complete unit cells each containing an A sublattice site and a B sublattice site (each unit cell is denoted by a dotted box), an intra-cell coupling of $t_1\pm\gamma$ and inter-cell coupling of $t_2$, and an additional A sublattice site at the right end at $n=N$ connected to the A sublattice site at the left end at $n=1$ by the terminal couplings $\lambda_{\mathrm{L}/\mathrm{R}}$. \textbf{b}. The complex eigenenergy spectrum of the chain in a. as a function of $N$ for $t_1=2$ and $t_2=1.5$. The blue ellipse at the top denotes the PBC eigenenergy spectrum of the corresponding homogenous periodic SSH chain. \textbf{c}. The logarithm of $|E_{\mathrm{min}}|$ (green), and $E_{\mathrm{min}}$ restricted to real values for the system in b. \textbf{d}. The complex eigenenergy spectrum of the chain in a. as a function of $N$ for $t_1=1.5$ and $t_2=2.8$ and the eigenenergy spectrum of the corresponding homogenous periodic SSH chain, and \textbf{e} the value and logarithm of $E_{\mathrm{min}}$. $\gamma=1$ and $\lambda_{\mathrm{L}}=\lambda_{\mathrm{R}} = 10^{-7}$ in b. -- e. }
\label{gFig1}
\end{figure}

Budich \textit{et al.} recently proposed a class of non-Hermitian sensors they dubbed ``non-Heritian topological sensors'' (NTOS) \cite{budich2020non}. The NTOS comprises a non-Hermitian Su-Schrieffer-Heeger (SSH) chain with its two ends connected to each other by a terminal coupling as shown schematically in Fig. \ref{gFig1}a. Under certain conditions, the system hosts an eigenstate with a eigenenergy close to zero (Fig. \ref{gFig1}b--e). Specifically, the value of this near-zero eigenenergy has a very large sensitivity to the terminal coupling, thus enabling the NTOS to be used as a sensor in which the measurement output is correlated to the terminal coupling strength. A TE realization of the NTOS was recently experimentally demonstrated \cite{yuan2023non}. In that study, the terminal coupling corresponds to the capacitance of variable capacitors with movable plates and the NTOS was used to measure the relative linear and angular positions of the capacitance plates in the capacitor. Another demonstration is a related system functioning as a sensitive ohmmeter in which the terminal coupling corresponds to the measured resistance  \cite{konye2023non}. Despite the exponential sensitivity of the NTOS, it faces the problem of the saturation of the nearly-zero energy eigenstate with the system size  beyond some critical system size (Fig. \ref{gFig1}b, c).  This  begs the fundamental question regarding NTOSs \cite{budich2020non, koch2022quantum, yuan2023non, konye2023non} : Why does the sensitivity of a NTOS eventually become saturated as the system size increases? From a practical standpoint, this saturation phenomenon poses a significant challenge to the implementation and scalability of NTOSs.
  
In this study, we aim to address this fundamental question and uncover the underlying mechanisms behind the saturation of the sensing characteristics in NTOSs. Specifically, we consider an NTOS that comprises a SSH chain with $(N-1)$ complete unit cells, each of which  comprises an A and a B sublattice site with a non-reciprocal intra-unit cell coupling $(t_1 \pm \gamma)$ and a reciprocal inter-unit cell coupling $t_2$. An additional A sublattice site is inserted at one end of the chain and is coupled to the other end via terminal couplings $\lambda_{\mathrm{L},\mathrm{R}}$ to form a closed ring. We take $t_1$, $t_2$, $\lambda_{\mathrm{L},\mathrm{R}}$ and $\gamma$ to be real. The system is described by the Hamiltonian   
\begin{align}
	H = & \sum_{n=1}^{N-1} |\mathrm{A}, n\rangle (t_1 + \gamma) \langle B, n| +  |\mathrm{B}, n\rangle (t_1 - \gamma) \langle \mathrm{A}, n | \nonumber \\
	+ & \sum_{n=1}^{N-1} t_2 \left( |\mathrm{A}, n+1\rangle \langle \mathrm{B}, n| + |\mathrm{B}, n\rangle\langle\mathrm{A}, n+1| \right) \nonumber \\
	+ & |\mathrm{A}, 1\rangle \lambda_{\mathrm{L}} \langle \mathrm{A}, N| + |\mathrm{A}, N\rangle \lambda_{\mathrm{R}} \langle \mathrm{A}, 1| \label{H0}
\end{align}
where the first A sublattice site on the left is located at unit cell $n=1$, and the last A sublattice site at unit cell $n=N$. Here, we assume that the magnitude of $\lambda_{\mathrm{L,R}}$ is much smaller than those of $t_1$ and $t_2$.

There are two possible scenarios for the variation of the eigenenergy with the smallest modulus, which we denote as $E_{\mathrm{min}}$, with the system size. (We focus on $E_{\mathrm{min}}$ and not eigenenergies with larger magnitudes because the other eigenenergies have only a polynomial rather than exponential dependence on the system size. $E_{\mathrm{min}}$ is physically significant because given that the total number of lattice sites is odd, there is always a broken unit cell regardless of the parameter choices. This results in an isolated lattice site with zero energy in the absence of the end perturbations $\lambda_{\mathrm{L},\mathrm{R}}$. When the end perturbation is applied, this zero-energy state shifts to a finite enery, i.e., $E_{\mathrm{min}}$. In a topolectrical realization of the system, $E_{\mathrm{min}}$ gives the dominant contribution to the impedance, which can serve as a readout for the terminal couipling strength.) In the first scenario exemplified by Fig. \ref{gFig1}b and \ref{gFig1}c, $E_{\mathrm{min}}$ is real at small system sizes and increases exponentially with the system size $N$ until the system reaches a critical size $N_{\mathrm{c}}$.  Beyond $N_{\mathrm{c}}$, the size dependence of $E_{\mathrm{min}}$ saturates, and $E_{\mathrm{min}}$ gains an imaginary part. As the system size approaches infinity, its energy spectrum approaches that of the corresponding periodic system under periodic boundary conditions (PBC) (i.e., one in which the additional $A$ site at $n=N$ and the terminal couplings $\lambda_{\mathrm{L},\mathrm{R}}$ are replaced by the inter-unit cell coupling $t_2$ so that the system becomes periodic). As shown in  Fig. \ref{gFig1}b, the PBC spectrum exhibits a point gap (PG) topology in which the locus of the complex eigenenergies forms a closed curve that encloses but does not touch the complex energy origin. For a different set of coupling parameters, we have the second scenario exemplified by Fig. \ref{gFig1}d and \ref{gFig1}e, in which $E_{\mathrm{min}}$ remains real and decreases exponentially with increasing system size. In fact, as we shall prove later on, $E_{\mathrm{min}}$ approaches zero as the system size approaches infinity. In this scenario, the eigenenergy spectrum of the chain exhibits a line gap (LG) topology in which the eigenenergy curves form two closed loops separated by either the or imaginary energy axis (the latter of which is shown in Fig. \ref{gFig1}d), and do not enclose the  origin. Furthermore, in the examples shown in Fig. \ref{gFig1}b -- \ref{gFig1}e, $E_{\mathrm{min}}$ fluctuates between positive and negative values with the variation of $N$. 

In this study, we analyze and explain the above behavior of the non-Hermitian SSH chain system. We first derive analytical expressions for the linear variation of $\mathrm{ln}\ |E_{\mathrm{min}}|$ with the system size (i.e., the linear regime). Subsequently, we explain why $|E_{\mathrm{min}}|$ decreases exponentially with the system size in the LG configuration but increases exponentially in the PG  topology up to a critical size. We derive the critical system size and energy at which the eigenenergy variation saturates in the PG configuration. Finally, we show that by introducing a unidirectional terminal coupling, i.e., setting either $\lambda_{\mathrm{L}}$ or $\lambda_{\mathrm{R}}$  to zero,  the behavior of the  system in the PG configuration can be switched, such that $E_{\mathrm{min}}$ decreases exponentially to zero instead of increasing as the system size increases.

\section{Linear regime}

\begin{figure}[ht!]
\centering
\includegraphics[width=0.48\textwidth]{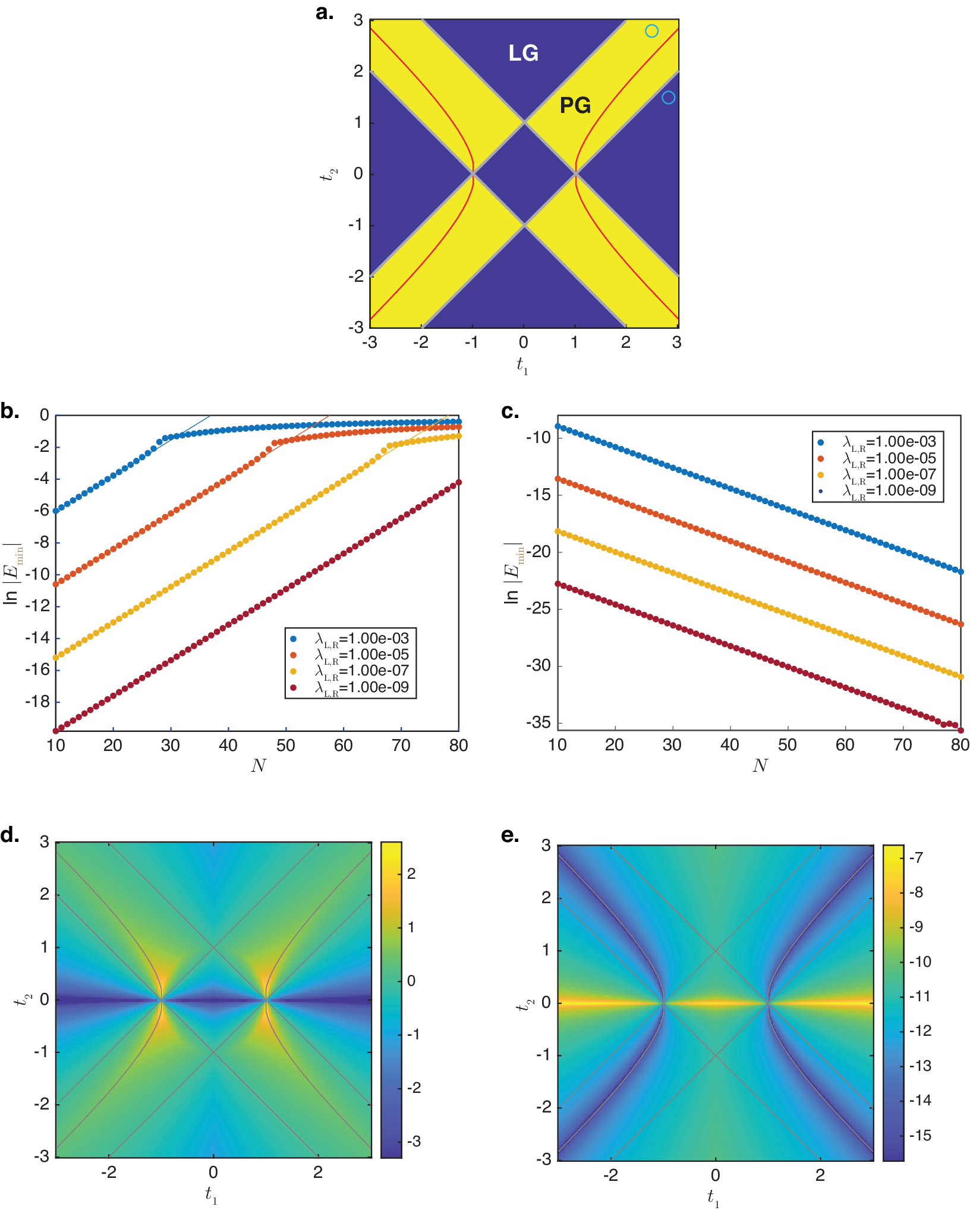}
\caption{\textbf{a}. $t_1$--$t_2$ plane phase diagram of the LG (blue) and PG (yellow) topologies at $\gamma=1$. The dark gray lines demarcating the boundaries between the LG and PG topologies correspond to $t_2 = \pm (t_1 + \gamma)$ and $t_2 = \pm (t_1 - \gamma)$ and the red lines $t_1^2 =  t_2^2-\gamma^2$. The two circles denote the $(t_1,t_2)$ coordinates of $(2.5,2.8)$ and $(2.8,1.5)$ shown in b. and c., respectively.  \textbf{b}.  $\mathrm{ln}\ |E_{\mathrm{min}}|$ obtained via direct numerical diagonalization (dotted lines) and calculated from Eq. \eqref{linE0} (continuous lines) for $(t_1,t_2)=(2.5,2.8)$ and $\gamma=1$ in the PG topology and the $\lambda_{\mathrm{L}}=\lambda_{\mathrm{R}}$ values of $10^{-3}$, $10^{-5}$, $10^{-7}$ and $10^{-9}$.  \textbf{c}. The counterpart to b. for the LG topology with $(t_1,t_2)=(2.8,1.5)$.  \textbf{d}. and \textbf{e}. The d. gradient and e. $\mathrm{ln}\ |E_{\mathrm{min}}|$ axis intercept of the variation of $\mathrm{ln}\ |E_{\mathrm{min}}|$ with $\mathrm{N}$ in Eq. \eqref{linE0} as functions of $t_1$ and $t_2$ for $\gamma=1$ and $\lambda_{\mathrm{L}} = \lambda_{\mathrm{R}} = 10^{-5}$. The  $|t_2| = |t_1 \pm \gamma|$ lines demarcating the boundaries between the LG and PG topologies and the $t_1^2 = t_2^2-\gamma^2$ lines demarcating the boundaries between the two signs of $s_t$ are indicated. }
\label{gFig2}
\end{figure}

\subsection{LG and PG topologies} 
We saw in Fig. \ref{gFig1} that the behavior of the NTOS differs depending on whether the PBC eigenenery spectrum of the corresponding homogeneous periodic system is in the LG or PG topology. To set the stage for further discussion, we first establish the conditions for the two topologies. Within the bulk of the NTOS (i.e., away from the $n=1, N$ unit cells at the two ends of the chain connected by the $\lambda_{\mathrm{L/R}}$ couplings), the eigenstates of the system are described by the non-Bloch Hamiltonian 
\begin{equation}
	H(\beta) = \begin{pmatrix} 0 & (t_1+\gamma) + t_2 / \beta \\ (t_1 - \gamma) + t_2 \beta & 0 \end{pmatrix}  \label{Hbeta}
\end{equation}
where $H(\beta)$ is written in the A, B sublattice basis. Because Eq. \eqref{Hbeta} is a 2 by 2 matrix, it admits two values of $\beta$ for each given eigenenergy value $E$. We denote the two values of $\beta$ as $\beta_1$ and $\beta_2$ where $|\beta_2|>|\beta_1|$ and the corresponding eigenvectors as $(1, \chi_j)^{\mathrm{t}}$, $j \in (1,2)$ where
\begin{equation}
	\chi_j = \frac{E \beta_j}{t_2 + \beta_j (t_1 + \gamma)} \label{chij}
\end{equation} 
and $\beta_{1,2}$ is one or the other of
\begin{widetext}
 \begin{equation}
	\beta_\pm = \frac{ E^2 - (t_1^2 + t_2^2 - \gamma^2) \pm \sqrt{ (t_1^2 - (E-t_2)^2-\gamma^2)(t_1^2 - (E+t_2)^2-\gamma^2)}}{2 t_2(t_1 + \gamma)} \label{betapm}
\end{equation}   
\end{widetext}
depending on the relative values of $t_1$, $t_2$, and $\gamma$. Because our main interest is on $E_{\mathrm{min}}$, i.e., the eigenenergy with the smallest value of $|E|$, we focus on small values of $|E|$, in which case $\beta_1 = \beta_-$ and $\beta_2 = \beta_+$ ($\beta_1 = \beta_+$ and $\beta_2 = \beta_-$ ) if $(t_1^2 - t_2^2 - \gamma^2) < 0$ ($(t_1^2 - t_2^2 - \gamma^2) > 0$). For later convenience, we also introduce the Taylor expansions of $\beta_\pm$ around $E=0$, $\beta_{\mathrm{a},\mathrm{b}}$ where
\begin{align}
	\beta_{\mathrm{a}} &= -\frac{t_1-\gamma}{t_2} + E^2  \frac{1}{t_1^2-t_2^2-\gamma^2} \label{betaa}, \\
	\beta_{\mathrm{b}} &= -\frac{t_2}{t_1+\gamma} - E^2 \frac{1}{t_1^2-t_2^2-\gamma^2} \label{betab}. 
\end{align}

The correspondence between $\beta_{a,b}$ and $\beta_{\pm}$ is listed in Table \ref{tab1}. (The entry for $t_1^2 + t_2^2 - \gamma^2 < 0$, $t_1^2 - t_2^2 - \gamma^2 > 0$ is empty because there are no real values of $t_1$, $t_2$, and $\gamma$ at which both $t_1^2 + t_2^2 - \gamma^2 < 0$ and $t_1^2 - t_2^2 - \gamma^2 > 0$ hold simultaneously. )

\begin{table}[]
	\caption{Correspondence between $\beta_\pm$, $\beta_{1,2}$, and $\beta_{\mathrm{a},\mathrm{b}}$}
	\begin{tabular}{|l|l|l|}
	\hline
	 & $(t_1^2 + t_2^2 - \gamma^2) > 0$   & $(t_1^2 + t_2^2 - \gamma^2) < 0$  \\ \hline
	 $(t_1^2 - t_2^2 - \gamma^2) > 0$ & \makecell{$\beta_1 = \beta_+, \beta_+ \approx \beta_{\mathrm{b}}$ \\ $\beta_2 = \beta_-, \beta_- \approx \beta_{\mathrm{a}}$}   & -  \\ \hline
	 $(t_1^2 - t_2^2 - \gamma^2) < 0$ & \makecell{$\beta_1=\beta_+, \beta_+\approx \beta_{\mathrm{a}}$ \\ $\beta_2=\beta_-, \beta_-\approx \beta_{\mathrm{b}}$}  & \makecell{$\beta_1=\beta_-, \beta_+ \approx \beta_{\mathrm{a}}$ \\ $\beta_2=\beta_+, \beta_- \approx \beta_{\mathrm{b}}$}  \\ \hline
	\end{tabular} \label{tab1}
\end{table}

The PG or LG topology of the corresponding PBC homogenous SSH system is quantified by a winding number $W$ defined as 
\begin{equation}
	W = \int^\pi_{-\pi} \frac{1}{2\pi i} \frac{\mathrm{d}}{\mathrm{d}k} \mathrm{ln} (\mathrm{det}\ H(k)) \label{W1},
\end{equation} 
where $\beta=\exp(ik)$. $|W|$ counts the number of times that the PBC eigenenergy curve of $H$ winds around the complex energy origin: $|W|=1$ in the PG topology because the PBC eigenenergy curve encloses the complex energy origin (Fig. \ref{gFig1}b) while $W=0$ in the LG topology because the PBC eigenenergy curve does not enclose the origin (Fig. \ref{gFig1}d) \cite{PRB109_045410}. From Eq. \eqref{Hbeta}, $W$ can be recast into the form of 
\begin{align}
	W =& -\frac{1}{2\pi i}\oint_{|\beta|=1} \mathrm{d}\beta\ \frac{\mathrm{d}}{\mathrm{d}\beta}  \mathrm{ln} \left( (t_1+\gamma)+t_2 \beta^{-1} \right) \nonumber \\
	&- \frac{1}{2\pi i}\oint_{|\beta|=1} \mathrm{d}\beta\ \frac{\mathrm{d}}{\mathrm{d}\beta}  \mathrm{ln} \left( (t_1-\gamma)+t_2 \beta \right).
\end{align}

The argument principle implies that the integral in the first row is 0 if $|t_2/(t_1+\gamma)| < 1$ and -1 otherwise, while that in the second row is 1 if $|(t_1-\gamma)/t_2| < 1$ and 0 otherwise. The system is therefore in the PG topology with $W=-1$ ($W=1$) if both $|t_2/(t_1+\gamma)|$ and $|(t_1-\gamma)/t_2|$ are simultaneously larger (smaller) than 1, and in the LG topology if one is bigger than 1 while the other is smaller. For the PG topology, the sign of $W$ indicates whether the open boundary condition (OBC) eigenstates on the generalized Brillouin zone (GBZ) in the thermodynamic limit of the system are localized towards the left ($W=1$) or right ($W=-1$) of the system \cite{PRB109_045410}.  The boundaries between the LG and PG topologies at which $|t_2/(t_1+\gamma)|=1$ and $|(t_1-\gamma)/t_2|=1$ are located at $t_2 = \pm (t_1+\gamma)$ and $t_2 = \pm (t_1-\gamma)$ on the $t_1$--$t_2$ plane and indicated as dark gray lines in Fig. \ref{gFig2}a. For  $|t_1|$ and $|t_2|$ larger than $|\gamma|$, the PG topology holds for values of $|t_2|$ between $|t_1-\gamma|$ and $|t_1+\gamma|$. 

We next investigate the dependence of the $N$ variation of $E_{\mathrm{min}}$ within the linear regime on the SSH chain parameters $t_1$, $t_2$, $\gamma$, and $\lambda_{\mathrm{L},\mathrm{R}}$.

\subsection{Eigenenergy variation within linear regime}

We derive an analytic expression for $E_{\mathrm{min}}$ as a function of $N$ in the linear regime. To do so, we first write the wavefunction of an eigenstate of Eq. \eqref{H0}, $\psi(n)$ where $n$ indexes the unit cell, as a linear combination of the eigenstates of Eq. \eqref{Hbeta}: 
\begin{equation}
	\psi(n) = \beta_1^n \begin{pmatrix} 1 \\ \chi_1 \end{pmatrix} + c \beta_2^n \begin {pmatrix} 1 \\ \chi_2 \end{pmatrix} \label{psin}
\end{equation}
where $c$ is the relative weight between the $\beta_1$ and $\beta_2$ eigenstates.  

For notational convenience, we introduce $\psi^{(\mathrm{A,B})} \equiv \langle \mathrm{A}, \mathrm{B} |\psi(n)$. From Eq. \eqref{H0}, the time-independent Schroedinger equation $H\psi(n) = E\psi(n)$ at $n=1$ and $n=N$ yields the boundary conditions 
\begin{align}
	(t_1+\gamma)\psi^{(\mathrm{B})}(1) + \lambda_{\mathrm{R}}\psi^{(\mathrm{A})}(N) &= E \psi^{(\mathrm{A})}(1) \label{bc1}, \\
	t_2\psi^{(\mathrm{B})}(N-1) + \lambda_{\mathrm{L}}\psi^{(\mathrm{A})}(1) &= E \psi^{(\mathrm{A})}(N) \label{bc2}. 
\end{align}
We substitute Eq. \eqref{psin} into Eq. \eqref{bc1} and \eqref{bc2}, and then substitute Eq. \eqref{chij} into the resulting expressions. From the requirement that the values of $c$ in the two expressions should match, we obtain the consistency condition

\begin{widetext}
  \begin{equation}
	\frac{ \beta_1 \left( (t_2 + \beta_1(t_1+\gamma))\lambda_{\mathrm{L}} - E\beta_1^N(t_1+\gamma)\right)}{ \beta_2 \left( (t_2 + \beta_2(t_1+\gamma))\lambda_{\mathrm{L}} - E\beta_2^N(t_1+\gamma)\right)} - \frac{ Et_2\beta_1 - \beta_1^N\left(t_2+\beta_1(t_1+\gamma)\right)\lambda_{\mathrm{R}}}{ Et_2\beta_2 - \beta_2^N\left(t_2+\beta_2(t_1+\gamma)\right)\lambda_{\mathrm{R}}} = 0.  \label{Deq0}
\end{equation}  
\end{widetext}

By applying a set of approximations on Eq. \eqref{Deq0} as described in detail in Appendix $\mathbf{A}$, the variation of $|E_{\mathrm{min}}|$ with $N$ in the linear regime can be obtained as
\begin{widetext}
   \begin{equation}
	\mathrm{ln}\ |E_{\mathrm{min}}|= N \mathrm{ln}\ \left|  \left(\frac{t_2}{t_1 + s_t s_g \gamma}\right)^{s_t}  \right| + \mathrm{ln} \left|\lambda_{s_g}\right|  + \left( \mathrm{ln}\ \left|  \frac{ t_1^2 - \gamma^2 - t_2^2 }{t_2 (t_1 - s_g \gamma)}  \right| \right) \label{linE0}
\end{equation} 
\end{widetext}

where 
\begin{align}
	s_t &\equiv \mathrm{sign} \left( \left| \frac{t_1-\gamma}{t_2} \right | - \left | \frac{t_2}{t_1+\gamma} \right| \right),  \label{st0} \\
	s_g &\equiv \mathrm{sign}(\mathrm{ln}\ \sqrt{\beta_1\beta_2}), \label{sg0} 
\end{align}
and 
\begin{equation}
	\lambda_{s_g} \equiv \begin{cases} 
		\lambda_{\mathrm{L}} & s_g < 0, \\
		\lambda_{\mathrm{R}} & s_g > 0
		\end{cases}. \label{lsg}
\end{equation}

$s_t$ reflects which of $\beta_\pm$ has a larger magnitude at $E=0$, while $s_g=1$ ($s_g=-1$) indicates that in the OBC thermodynamic limit of the homogeneous chain, the GBZ states are localized at the right edge (left) of the system. Notice that in the PG topology, $s_g$ is the negative of the winding number defined in Eq. \eqref{W1} where $s_g = -1$ and $W = 1$ signify GBZ states localized towards the left of the system while $s_g = 1$ and $W = -1$ signify GBZ states localized towards the right.

A few conclusions can be drawn from Eq. \eqref{linE0}. The coefficient of $N$ in Eq. \eqref{linE0}, i.e., the gradient of the $\mathrm{ln}\ |E_{\mathrm{min}}|$ versus $N$ line, is positive in the PG topology and negative in the LG topology. This can be seen by noting that the gradient becomes zero at $t_2 = \pm (t_1 + \gamma)$ and $t_2 = \pm (t_1 - \gamma)$,  which are precisely the boundary lines between the LG and PG topologies.  The sign of the gradient within each region bounded by the boundary lines can be then obtained by, for example, considering the limit of $t_1 \rightarrow \infty$. Therefore,  $|E_{\mathrm{min}}|$  increases exponentially with increasing $N$ in the PG topology (Fig. \ref{gFig1}b and \ref{gFig1}c) and decreases exponentially with increasing $N$ in the LG topology (Fig. \ref{gFig1}d and \ref{gFig1}e).

It can also be seen from Eq. \eqref{linE0} that the gradient is independent of $|\lambda_{s_g}|$ whereas the value of  $\mathrm{ln}\ |E_{\mathrm{min}}|$ in the $N=0$ limit varies linearly with $\mathrm{ln}\ |\lambda_{s_g}|$. These analytical predictions are borne out in by the close agreement in Fig. \ref{gFig2}b and \ref{gFig2}c between the numerical plot of $\mathrm{ln}\ |E_{\mathrm{min}}|$ calculated by the direct numerical diagonalization of the Hamiltonian Eq. \eqref{H0} and the analytical plot lines obtained using Eq. \eqref{linE0}. 

Fig. \ref{gFig2}d and \ref{gFig2}e show the $t_1$ and $t_2$ variation of the gradient and $\mathrm{ln}\ |E_{\mathrm{min}}|$-axis intercept of the $\mathrm{ln}\ |E_{\mathrm{min}}|$ versus $N$ graph at $\gamma=1$, respectively. (The same quantitative trends are observed at negative real values of $\gamma$.)  Within the PG topology, the gradient and $\mathrm{ln}\ |E_{\mathrm{min}}|$ axis intercepts have the largest magnitudes around the $t_1^2=t_2^2\pm \gamma^2$ lines. A large gradient implies a high sensitivity of $|E_{\mathrm{min}}|$ to the system size, whereas a highly negative $\mathrm{ln}\ |E_{\mathrm{min}}|$ intercept implies a larger range of $|E_{\mathrm{min}}|$ over which $|E_{\mathrm{min}}|$ varies exponentially with $N$. Both of these properties are desirable in the operation of the the SSH chain as a NTOS. The gradient in the PG topology decreases slowly as $|t_1|$ and $|t_2|$ increase whereas the $\mathrm{ln}\ |E_{\mathrm{min}}|$-axis intercept becomes more negative.  In comparison, for the LG topology, the gradient is the steepest and the $\mathrm{ln}\ |E_{\mathrm{min}}|$ intercept the largest around $t_2=0$. The latter implies that the largest values of $\mathrm{ln}\ |E_{\mathrm{min}}|$ for a given value of $N$ and $\lambda_{\mathrm{L,R}}$ occur around $t_2=0$. 

The plots of $\mathrm{ln}\ |E_{\mathrm{min}}|$ against $N$ for the LG topology in Fig. \ref{gFig1}e and \ref{gFig2}c do not preclude the possibility that the variation of $\mathrm{ln}\ |E_{\mathrm{min}}|$ with $N$ may saturate at some large value of $N$. As derived in the Appendix $\mathbf{B}$, the necessary condition for  $|E_{\mathrm{min}}|$  to tend to 0 as $N$ tends to infinity is either 
\begin{widetext}
  \begin{equation}
	\frac{\lambda_{\mathrm{L}}}{t_1+\gamma}\left(\frac{-t_1+\gamma}{t_2}\right)^N + \frac{\lambda_{\mathrm{R}}}{t_1-\gamma}\left( -\frac{t_2}{t_1+\gamma}\right)^{-N} \rightarrow 0\ \text{as}\ N\rightarrow \infty \label{c00}
\end{equation}	  
\end{widetext}

or
\begin{widetext}
  \begin{equation}
	\frac{\lambda_{\mathrm{L}}}{t_1+\gamma}\left(\frac{-t_1+\gamma}{t_2}\right)^{-N} + \frac{\lambda_{\mathrm{R}}}{t_1-\gamma}\left( -\frac{t_2}{t_1+\gamma}\right)^{N} \rightarrow 0\ \text{as}\ N\rightarrow \infty. \label{c00a}
\end{equation}  
\end{widetext}
	
(There are two alternatives because there are two possible choices for associating $\beta_{\mathrm{a}}$ and $\beta_{\mathrm{b}}$ with $\beta_1$ and $\beta_2$ in Eq. \eqref{Deq0}. ) 

For either of these conditions to be satisfied so that $E_{\mathrm{min}} \rightarrow 0$ as $N \rightarrow\infty$, we require one of $|\beta_{\mathrm{a}}^{(0)}|$ and $|\beta_{\mathrm{a}}^{(0)}|$ to be smaller than 1, and the other to be larger than 1 where $\beta_{\mathrm{a,b}}^{(0)}$ is the value of $\beta_{\mathrm{a,b}}$ at $E=0$. This is precisely the condition for the LG topology with a winding number of $W=0$. Therefore, $\mathrm{ln}\ |E_{\mathrm{min}}|$ approaches $-\infty$ without saturation in the LG geometry as $N \rightarrow \infty$. In contrast, because  only one of the two conditions holds in the PG topology (where $W=1$ [$W=-1$] corresponds to both $|\beta_a^{(0)}|$ and $|\beta_b^{(0)}|$ being larger [smaller] than 1), $|E_{\mathrm{min}}|$ does not decrease with increasing $N$ in the PG topology but instead increases until it reaches a critical value $E_{\mathrm{c}}$, which we derive in the next subsection. The magnitude of the winding number $|W|$ therefore indicates whether $|E_{\min}|$ increases ($|W|=1$ in the PG topology) or decreases ($W=0$ in the LG topology) as $N\rightarrow\infty$. 

The differences in the behavior of $E_{\mathrm{min}}$ between the LG and PG topologies can be further understood in the following way: Let us first consider the PG topology where, for definiteness, we consider the case that $1< |\beta_1|<|\beta_2|$, i.e., $W=1$. (The argument for the $W=-1$ case follows analogously.)  In this case, the difference between the magnitudes of $\psi^{(A)}(N)$ and $\psi^{(A)}(1)$ in Eq. \eqref{bc1} grows exponentially with $N$ because the magnitudes of both the $\beta_1^n$ and $\beta_2^n$ components in $\psi(n)$  increase with $N$. In order for Eq. \eqref{bc1} to still hold as $N$ increases, $|E|$ on the right-hand side of Eq. \eqref{bc1}  has to increase to compensate for the larger value of $|\psi^{(A)}(N)|$. However, $|E|$ cannot increase indefinitely because as $|E|$ increases, the disparity between $|\beta_1|$ and $|\beta_2|$ increases as well. (This can be seen by noting that the lowest power in $E$, i.e., $E^2$, of $\beta_{\mathrm{a}}/\beta_{\mathrm{b}}$ from Eq. \eqref{betaa} and \eqref{betab} has a positive coefficient of $(t_1^2-t_2^2-\gamma^2)^{-2}$. ) This implies that there exists some critical value of $|E|$, beyond which the left-hand side of Eq. \eqref{bc1} starts to grow at a faster rate than can be compensated for by the increase in $E\psi^{(A)}(1)$ on the right-hand side. We derive this critical value of $|E|$ and the system size at which it occurs, i.e., $N_{\mathrm{c}}$,  in the next section.   

By contrast, in the LG topology where $|\beta_1|<1<|\beta_2|$ , the $\beta_1$ and $\beta_2$ components in the constituent wavefunction grow exponentially in opposite directions along the chain so that the relative magnitudes of the wavefunction $|\psi(1)|$ and $|\psi(N)|$ at the two ends can be largely kept relatively constant. This avoids the abovementioned scenario  where the difference between the magnitudes of $\psi(1)$ and $\psi(N)$ at the two ends grows indefinitely large as $N$ increases. Hence, no saturation of $E$ occurs in the LG topology.

Nonetheless, we can show that the eigenenergy spectrum of the bulk states approaches the PBC spectrum of the corresponding homogenous chain regardless of whether the NTOS is in the LG or PG topology (the detailed derivations are given in Appendix $\mathbf{B}$). This behavior is reminiscent of that exhibited by SSH chains without the additional A site at the end, i.e., when the A sublattice site at one end is connected directly to the B sublattice site at the other end by a small terminal coupling (thus constituting a different system from the NTOS studied here where the $\lambda_{\mathrm{L},\mathrm{R}}$ couplings is between A sublattice sites at the two ends) which were studied in Refs. \cite{guo2021exact} and \cite{SciRep13_22770}. In the system studied in Refs. \cite{guo2021exact} and \cite{SciRep13_22770}, the terminal coupling act as defect sites because the couplings to their immediate neighbors differ from those of the lattice sites in the interior of the chain. The effect of these defect sites become less prominent as the system length increases because the system can support longer-wavelength modes, which are less sensitive to presence of the highly localized defect sites. We can argue that a similar phenomenon is occurring in the present NTOS system where the extra A sublattice site plays the role of the defect site.

\subsection{Saturation of energy variation with system size $N$ in PG topology}

\begin{figure}[htp]
\centering
\includegraphics[width=0.48\textwidth]{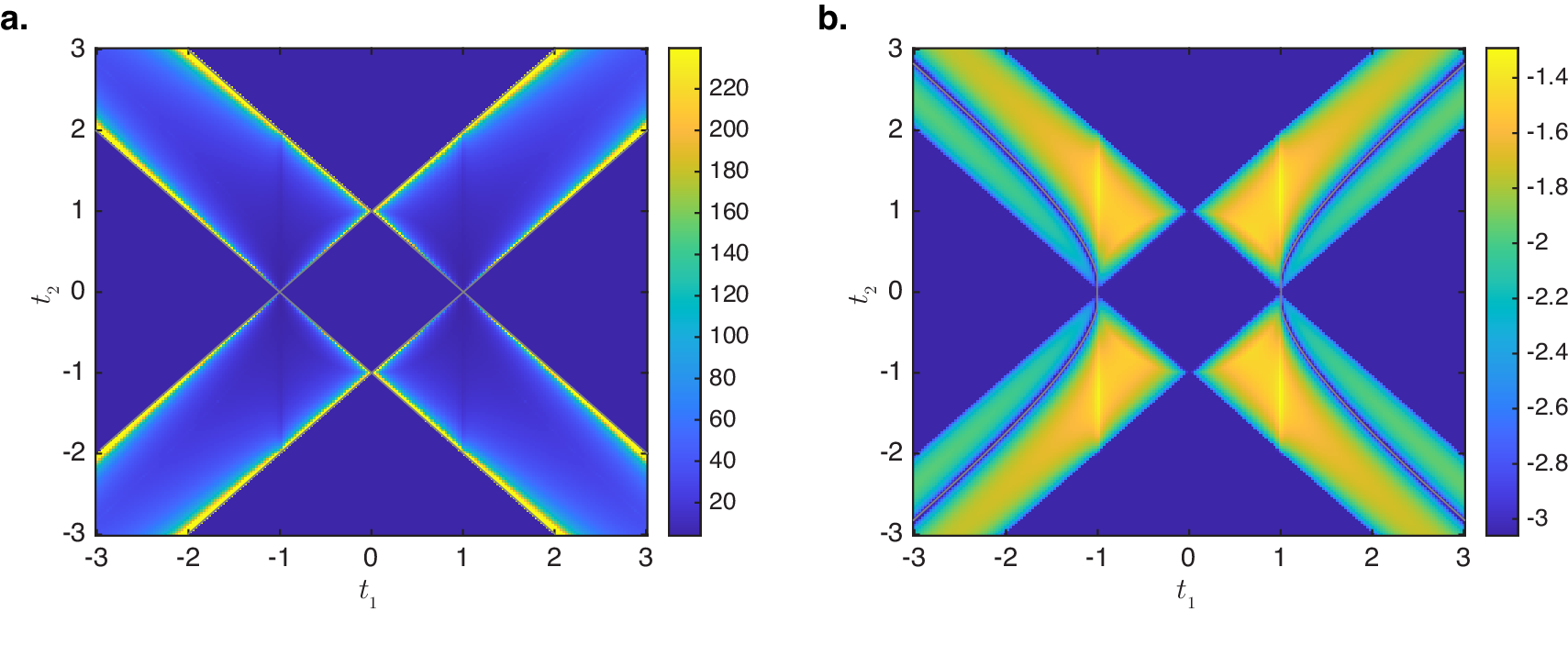}
\caption{  \textbf{a}. $N_{\mathrm{c}}$ and \textbf{b}. $\mathrm{ln}\ |E_{\mathrm{c}}|$ plotted as functions of $t_1$ and $t_2$  for $\gamma=1$ and $\lambda_{\mathrm{L}} = \lambda_{\mathrm{R}} = 10^{-7}$. }
\label{gFig3}
\end{figure}	

In the previous section, we noted that in the PG topology, $|E_{\mathrm{min}}|$ increases exponentially with $N$ before saturating at some critical system size $N_{\mathrm{c}}$. The critical size $N_{\mathrm{c}}$ and the energy $E_{\mathrm{c}}$ at which the saturation occurs can be approximately derived (see Appendix $\mathbf{C}$ for the detailed derivations) and are given by
\begin{widetext}
    \begin{align}
	E_{\mathrm{c}}^{(|\beta|<1)} &\approx \frac{ \sqrt{ \mathrm{ln} \left| \left(\frac{s_t \gamma - t_1}{t_2}\right)^{s_t} \right| }} { \sqrt{ - \left| \frac{1}{t_1^2-t_2^2-\gamma^2} \right| }\sqrt{\mathcal{W}\left( \frac{ t_2^2(t_1-\gamma)^2\left|t_1^2-t_2^2 - \gamma^2\right|\mathrm{ln} \left| \left(\frac{s_t \gamma - t_1}{t_2}\right)^{s_t} \right|}{-\mathrm{e}(t_1^2-t_2^2+\gamma^2)^2\lambda_{\mathrm{L}}^2}\right)}}, \label{E0x2} \\
    N_{\mathrm{c}}^{(|\beta|<1)} &\approx 2 + \left| \frac{t_1^2-t_2^2-\gamma^2}{2 \left(E_{\mathrm{c}}^{(|\beta|<1)|}\right)^2} \right| \label{M0x1}
\end{align}
\end{widetext}

when $|\beta_1|<|\beta_2|<1$,  and by 
\begin{widetext}
\begin{align}
    E_{\mathrm{c}}^{(|\beta|>1)} &= \frac{ \sqrt{ \mathrm{ln} \left| \left(\frac{s_t \gamma + t_1}{t_2}\right)^{-s_t} \right| }} { \sqrt{  \left| \frac{1}{t_1^2-t_2^2-\gamma^2} \right| }\sqrt{\mathcal{W}\left(\left| \frac{ (t_1^2-\gamma^2)^2\mathrm{ln} \left| \left(\frac{s_t \gamma + t_1}{t_2}\right)^{-s_t} \right|}{-\mathrm{e}(t_1^2-t_2^2-\gamma^2)\lambda_{\mathrm{R}}^2}\right|\right)}} \label{E0x3} \\
    N_{\mathrm{c}}^{(|\beta|>1)} &= 1 + \left| \frac{t_1^2-t_2^2-\gamma^2}{2 (E_{\mathrm{c}}^{(|\beta|>10|)})^2} \right| \label{M0x3}
\end{align}
\end{widetext}
when $1 < |\beta_1| < |\beta_2|$ where $\mathcal{W}(z)$ is the Lambert $W$ function, which gives the solution to $w$ for $w\exp(w)=z$.

Fig. \ref{gFig3}a and  \ref{gFig3}b respectively show the values of $N_{\mathrm{c}}$ and $\mathrm{ln}\ |E_{\mathrm{c}}|$ on the $t_1$--$t_2$ plane calculated using the analytical expressions in Eqs. \eqref{E0x2}, \eqref{M0x1}, \eqref{E0x3}, and \eqref{M0x3}. Fig. \ref{gFig3}a shows that the peak values of $N_{\mathrm{c}}$ occur at the boundaries between the LG and PG topologies. This may be explained by noting from Fig. \ref{gFig2}d that the boundaries between the LG and PG topologies correspond to the regions on the $t_1$--$t_2$ plane at which the gradient of the $\mathrm{ln}\ |E_{\mathrm{min}}|$ versus $N$ curve has the smallest absolute values, i.e., the regions where $|E_{\mathrm{min}}|$ increases most slowly with $N$. Thus, $N$ can reach a large value before $|E_{\mathrm{min}}|$ saturates. The gradual increase in $N_{\mathrm{c}}$ with the simultaneous increase of $|t_1|$ and $|t_2|$ (i.e., in the vicinity of the $|t_1| = \pm |t_2|$ directions) in Fig. \ref{gFig3}a is also consistent with the slow decrease along the same directions in the gradient $d(\mathrm{ln}\ |E_{\mathrm{min}}|)/dN$ in the linear regime shown in Fig. \ref{gFig2}d (the gradients have smaller values further away from $(t_1,t_2) = 0$. ). The values of $N_{\mathrm{c}}$ in the vicinity of the $t_1^2 = \pm t_2^2+\gamma^2$ lines are comparatively smaller.  Fig. \ref{gFig3}b clearly shows that the minimum values of $|E_{\mathrm{c}}|$ occur in the vicinity of the $t_1^2 = \pm t_2^2+\gamma^2$ lines. The minimum values of $|E_{\mathrm{c}}|$  there are consistent with the large negative intercepts of the $\mathrm{ln}\ |E|$ versus $N$ curves shown in Fig. \ref{gFig2}e. These results indicate that for practical applications of the SSH chain as an NTOS, the advantage of high sensitivity of $|E_{\mathrm{min}}|$ to $N$ in the vicinity of the $t_1^2 = \pm t_2^2+\gamma^2$ lines is partially offset by the small peak values of $N$ and $E_{\mathrm{min}}$ before the onset of saturation.

\section{Unidirectional coupling} 
\begin{figure}[htp]
\centering
\includegraphics[width=0.48\textwidth]{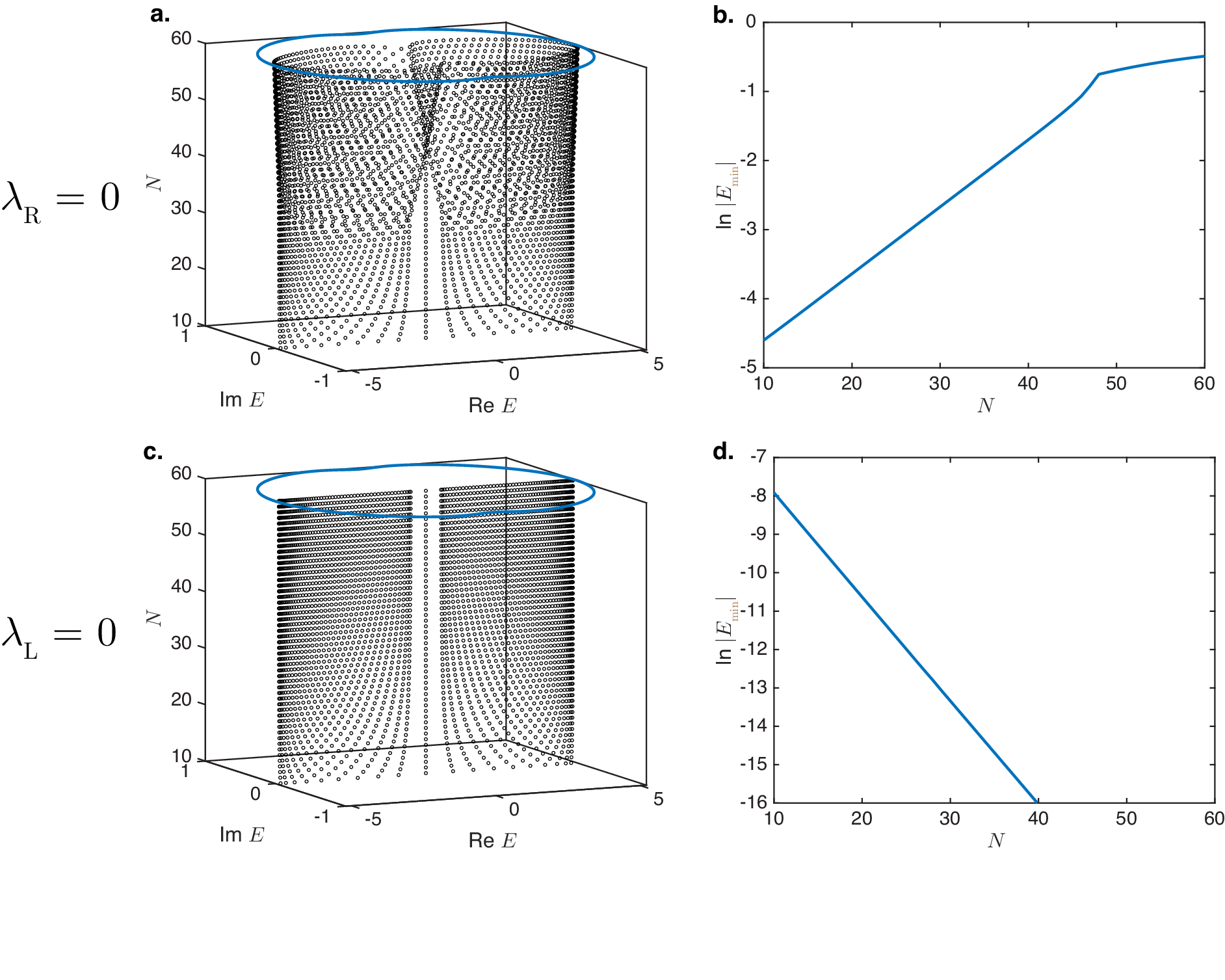}
\caption{\textbf{a}. Variation of complex eigenenergy spectrum with $N$ and the PBC spectrum of the corresponding homogeneous periodic chain for a SSH system with $t_1=2.5$, $t_2=2.8$, $\gamma=1$, $\lambda_{\mathrm{L}} = 10^{-5}$, and $\lambda_{\mathrm{R}} = 0$. \textbf{b}.  Variation of $\mathrm{ln}\ |E_{\mathrm{min}}|$ with $N$ for system in a.  \textbf{c} Variation of complex eigenenergy spectrum with $N$ for same system in a. except that $\lambda_{\mathrm{L}} = 0$, $\lambda_{\mathrm{R}} = 10^{-5}$. \textbf{d}.  Variation of $\mathrm{ln}\ |E_{\mathrm{min}}|$ with $N$ for system in c.}
\label{gFig4}
\end{figure}	

We now consider the unidirectional coupling case, in which either $\lambda_{\mathrm{L}}$ or $\lambda_{\mathrm{R}}$ is set to zero. Interestingly, the size-dependence of the unidirectional system in the PG topology can be switched by setting either $\lambda_{\mathrm{L}}$ or $\lambda_{\mathrm{R}}$ to zero. Fig. \ref{gFig4} shows the eigenspectrum and variation of $\mathrm{ln}\ |E_{\mathrm{min}}|$ with $N$ for a system with PG symmetry and having unidirectional coupling (i.e., either  $\lambda_{\mathrm{L}} = 0$ or $\lambda_{\mathrm{R}}=0$ ), with all other system parameters being the same as those in Fig. \ref{gFig1}b. In the case where  $\lambda_{\mathrm{R}}=0$ (see Fig. \ref{gFig4}b), the behavior of the resultant system is very similar to that of the original system in Fig. \ref{gFig1}c, i.e., $|E_{\mathrm{min}}|$ increases exponentially with the system size $N$  up to a critical value while the eigenenergy spectrum of the other states tends towards the PBC spectrum of the corresponding homogeneous system (Fig. \ref{gFig4}a). In contrast, the behavior of the system changes dramatically when $\lambda_{\mathrm{L}}$ rather than $\lambda_{\mathrm{R}}$ is set to 0. $|E_{\mathrm{min}}|$ now decreases to 0 as $N\rightarrow\infty$ (Fig. \ref{gFig4}d) while the eigenenergy spectrum of the remaining states tends towards the GBZ of the OBC spectrum rather than PBC spectrum of the corresponding homogenous system (Fig. \ref{gFig4}c). 

We will now analytically explain the switch in the size-dependence of the system between the $\lambda_{\mathrm{L}} = 0$ and  $\lambda_{\mathrm{R}} = 0$ cases. We define  the variables $s_{g'}$ and $\lambda_{s_{g'}}$ as follows: we set $s_{g'} = -1$ and $\lambda_{s_{g'}}=\lambda_{\mathrm{L}}$ when $|\lambda_{\mathrm{L}}|>0$ and $\lambda_{\mathrm{R}}=0$, and set $s_{g'} = 1$ and $\lambda_{s_{g'}}=\lambda_{\mathrm{R}}$ when $|\lambda_{\mathrm{R}}|>0$ and $\lambda_{\mathrm{L}}=0$. In the derivation of Eq. \eqref{linE0} in the consistency condition obtained from the boundary conditions Eq. \eqref{bc1} and \eqref{bc2}, we consider the effect of the term with the dominant influence between $\lambda_{\mathrm{L}}$, $\lambda_{\mathrm{R}}$, which is decided by $s_g$, and neglected the influence of the other term  (refer to Appendix $\mathbf{A}$, Eqs. \eqref{D0} to \eqref{e35}).  


For the $\lambda_{\mathrm{R}} = 0$ case, $s_{g'} = s_g$ as defined in Eq. \eqref{sg0} for the $t_1$, $t_2$, and $\gamma$ values in Fig. \ref{gFig4}. Hence, the analytical derivation of $\mathrm{ln}\ |E_{\mathrm{min}}|$ as a function of $N$ proceeds in an identical fashion to that leading to Eq. \eqref{linE0}, the corresponding equation for the non-unidirectional case i.e., with $\lambda_{\mathrm{R}} \neq 0$. Hence, the variation of $|E_{\mathrm{min}}|$ shows the same upward trend with $N$ as before, as shown in Fig. \ref{gFig4}a and b even when unidirectional coupling is imposed.


In contrast, for the  $\lambda_{\mathrm{L}} = 0$ case, the ``dominant'' term in Eq. \eqref{D0} is the second term, i.e., $\lambda_{\mathrm{R}} e^{g(N-1)}$  since the first term goes to zero when$\lambda_{\mathrm{L}} = 0$.  Eq. \eqref{linE0} would then need to be modified by replacing $s_g$ with $s_{g'}=-s_g$.  Note that because $| t_2/(t_1 \mp \gamma) | < 1 < | t_2/(t_1 \pm \gamma) |$ in the PG topology (see, e.g., Fig. \ref{gFig2}a), the sign of the coefficient of $N$ in Eq. \eqref{linE0} switches from positive to negative when $s_g$ is replaced by $s_{g'}=-s_g$. $|E_{\mathrm{min}}|$ therefore decreases with increasing $N$, as shown in Fig. \ref{gFig4}c and d.  (This is in contrast to a system with the LG topology where $\mathrm{ln}\ |t_2/(t_1 \pm \gamma) |$ have the same signs, such that the sign of the coefficient of $N$ remains unchanged  when $s_g$ is replaced by $-s_g$.) 

 
We now consider the conditions for $|E_{\mathrm{min}}|$ to approach zero as $N \rightarrow \infty$. As mentioned earlier, we require either $|\beta_a^{(0)}| < 1$ and $|\beta_b^{(0)}| > 1$, or $|\beta_a^{(0)}| > 1$ and $|\beta_b^{(0)}| < 1$  so that either Eq. \eqref{c00} or \eqref{c00a} is satisfied . As we discussed earlier, both of these conditions cannot be met when the system has the PG topology. However, when either $\lambda_{\mathrm{L}}$ or $\lambda_{\mathrm{R}}$ is 0, the conditions are relaxed such that one of Eq. \eqref{c00} or \eqref{c00a} will always be satisfied as $N \rightarrow \infty$. (E.g., when $\lambda_{\mathrm{R}} = 0$ and $\beta_a = (-t_1+\gamma)/t_2$, Eq. \eqref{c00} is satisfied if $|\beta_{\mathrm{a}}^{(0)}| < 1$ and Eq. \eqref{c00a} satisfied if $|\beta_{\mathrm{a}}^{(0)}| > 1$.). It now becomes possible for $|E_{\mathrm{min}}|$ to approach $0$ in the thermodynamic limit even in the PG topology. This possibility for $|E_{\mathrm{min}}|\rightarrow 0$ as $N\rightarrow \infty$  can be further understood as follows: Earlier when we considered a PG system with bidirectional coupling where $1 < |\beta_1| < |\beta_2|$, it was argued that $|E|$ on the right-hand side of Eq. \eqref{bc1} has to grow with increasing $N$ in order to compensate for the larger magnitude of $\lambda_{\mathrm{R}} \psi^{(A)}(N)$ as $N$ increases. However, when $\lambda_{\mathrm{R}}$ is now set to 0, this mechanism that drives the increase of $|E|$ with $N$ is no longer applicable in the PG system. 

We finally show that bulk energy spectrum approaches the OBC limit when unidirectional coupling is imposed on the system such that $s_{g'}\neq s_g$, i.e., when the direction of the finite unidirectional coupling is mismatched with the value of $s_g$. For definiteness, consider the $\lambda_{\mathrm{L}}=0$ case (i.e., $s_{g'}=1$) depicted in Figs. \ref{gFig4}c and \ref{gFig4}d, for which the values of $t_1$, $t_2$, and $\gamma$ correspond to $s_{g}=1$ so that $s_{g} = -s_{g'}$. The consistency condition Eq. \eqref{Deq0} reduces to 
\begin{equation}
		E \left(\beta_1^{-N} - \beta_2^{-N}\right) + \lambda_{\mathrm{R}} \left(\frac{1}{\beta_1} - \frac{1}{\beta_2}\right) = 0. \label{uniBulk0}
\end{equation}

When $|\beta_1|<1$ and $|\beta_2|<1$ in the vicinity of the GBZ, the first term in Eq. \eqref{uniBulk0} dominates in the limit of large $N$. Thus, Eq. \eqref{uniBulk0} is then approximately satisfied by $\beta_2 = \beta_1 \exp(i (2\pi m/N))$ where $m$ is an integer.  $\beta_1$ and $\beta_2$ then have the same moduli, or in other words, the bulk eigenenergy values lie on the OBC complex energy plane GBZ, which corresponds to the set of energy eigenvalues at which $|\beta_1|=|\beta_2|$ . The large-$N$ limit of the bulk energy spectrum when $\lambda_{\mathrm{L}} = 0$ is therefore the OBC GBZ when $|\beta|<1$ on the GBZ. The $|\beta|$ values on the GBZ are given by $\exp(g)$ given in Eq. \eqref{expg}, which is less than 1 when $t_1$ and $\gamma$ have the same signs, or in other words, when $s_g = -1 = -{s_g'}$,. Therefore, when $s_g = -s_{g'}$, the eigenenergy spectrum of the bulk states approaches the OBC GBZ. 

In contrast, when it is $\lambda_{\mathrm{L}}$ rather than  $\lambda_{\mathrm{R}}$ that has a finite value and $|\beta|<1$ on the GBZ, $s_{g}=s_{g'}=-1$. The finite $\lambda_{\mathrm{L}}$ analogue of the arguments following Eq. \eqref{bulkX1} and \eqref{bulkX2} in the Appendix $\mathbf{B}$, which consider only the finite $\lambda_{L}$ term, then apply and the eigenenergy spectrum of the system tends towards the PBC spectrum.  


\section{Conclusion}
In conclusion, we investigated the sensing characteristics of a NTOS and address gaps in the fundamental understanding of its behaviour in terms of its topology (LG - line gap or PG - point gap) and directionality of its terminal coupling (bidirectional or unidirectional).

We focus on the minimum eigenstate $E_{\mathrm{min}}$, which plays a key role in the impedance readout of a TE circuit realization of the NTOS. We showed that the system size dependence of the NTOS depends critically on whether the system is in the LG or PG topology. In the PG topology, $E_{\mathrm{min}}$ grows exponentially with increasing system size up to a critical size in systems with bidirectional coupling. We analytically derived the coupling relation which corresponds to the largest gradient of increase of $E_{\mathrm{min}}$ with respect to $N$, i.e., when $t_1^2=t_2^2\pm \gamma^2$. Such an exponential increase of $|E_{\mathrm{min}}|$ in the PG topology may be utilized directly to amplify the NTOS voltage readout. In contrast, in the LG topology, $|E_{\mathrm{min}}|$ decreases exponentially to zero with increasing system size, with the largest gradient of decrease in $|E_{\mathrm{min}}|$ with respect to $N$ occurring near$ t_2 = 0$. In a TE realization of the NTOS, such an exponential decrease in $|E_{\mathrm{min}}|$ can still translate to a large sensor readout, by converting it to the impedance of the circuit, which is inversely proportional to $|E_{\mathrm{min}}|$ \cite{PRB107_245114}.  

We also elucidated the phenomenon in which $|E_{\mathrm{min}}|$ saturates when the system size is increased beyond a certain critical size and derived the critical system size $N_{\mathrm{c}}$ and energy $E_{\mathrm{c}}$ at which the saturation occurs. This saturation phenomenon is important from the practical standpoint as the advantage of high sensitivity of $|E_{\mathrm{min}}|$ to $N$ would be offset by small values of $N_{\mathrm{c}}$ and $E_{\mathrm{c}}$ at saturation.

We finally investigated the effect of unidirectional  terminal couplings on the NTOS sensing behavior. Interestingly, we found that the exponential increase of $|E_{\mathrm{min}}|$ with the system size $N$ in the PG topology can be switched to an exponential decrease with $N$ by reversing the unidirectional direction of the terminal coupling. In summary, our study highlights and explains the importance of the PBC eigenspectrum topology, system size, and directionality of the terminal couplings in the design and optimization of non-Hermitian topological sensors.

\section{Appendix}

\subsection{Appendix A: Derivation of Eqs. \eqref{linE0}, \eqref{c00}, and \eqref{c00a} for linear regime}

For convenience, we write $\beta_1 = \exp(g)\exp(-i\theta)$ and $\beta_2=\exp(g)\exp(i\theta)$ where $\exp(g) = \sqrt{\beta_1\beta_2}$. From Eq. \eqref{betapm}, this gives
\begin{align}
	\exp(g) &= \sqrt{\frac{ t_1-\gamma}{t_1+\gamma}}, \label{expg} \\
	\exp(i\theta) &= \sqrt{ \frac{\beta_2}{\beta_1} }. \label{expitheta} 
\end{align}

$\exp(g)$ is the value of $|\beta_1|=|\beta_2|$ on the GBZ of the homogenous system in the thermodynamic limit when OBCs are imposed. Note that in general, $\theta$ in Eq. \eqref{expitheta} is complex and $|\exp(i\theta)| \neq 1$ unless $E$ lies on the complex energy plane GBZ of the homogenous system. 
 
Labeling the quantity on the left-hand side of Eq. \eqref{Deq0} as $D$, Eq. \eqref{Deq0} can be written as  
\begin{align} 
	D =&  -\frac{E (\mathrm{e}^{-g(N-1)}\lambda_{\mathrm{L}} + \mathrm{e}^{g(N-1)}\lambda_{\mathrm{R}} )}{E^2 - \lambda_{\mathrm{L}}\lambda_{\mathrm{R}} } \sin(\theta) \nonumber \\
	& -\frac{\lambda_{\mathrm{L}}\lambda_{\mathrm{R}}}{E^2 - \lambda_{\mathrm{L}}\lambda_{\mathrm{R}}}\sin((N-2)\theta) \nonumber \\ 
	& -\left( \frac{t_2}{\mathrm{e}^g(t_1+\gamma)} + \frac{\mathrm{e}^g(t_1+\gamma)}{t_2}\right)\frac{\lambda_{\mathrm{L}}\lambda_{\mathrm{R}}}{E^2-\lambda_{\mathrm{L}}\lambda_{\mathrm{R}}}\sin((N-1)\theta) \nonumber\\
	& +\sin(N\theta) = 0\label{Dq0}.
\end{align}

Introducing 
\begin{equation}
	s_t = \mathrm{sign} \left( \left| \frac{t_1-\gamma}{t_2} \right | - \left | \frac{t_2}{t_1+\gamma} \right| \right), \label{st}
\end{equation} 
we have, from Eqs. \eqref{betaa} and \eqref{betab},
\begin{align}
	\beta_2^{(0)} = -\left(\frac{t_1-s_t\gamma}{t_2}\right)^{s_t} \label{beta20} \\
	\exp(-i\theta^{(0)}) = - \left (\frac{t_2}{\sqrt{t_1^2-\gamma^2}} \right)^{s_t} \label{eiq0}
\end{align}
where $\beta_2^{(0)}$ and $\exp(-i\theta^{(0)})$ are the values of $\beta_2$ and $\exp(i\theta)$ at $E=0$. From its definition in Eq. \eqref{st}, it can be seen that $s_t$ changes sign at $t_2^2 = \pm(t_1^2-\gamma^2)$, which is indicated by the red lines in Fig. \ref{gFig2}a.

The numerical results in Fig. \ref{gFig1}c and e indicate that $E_{\mathrm{min}}$ satisfy $|E_{\mathrm{min}}| \gg |\lambda_{L,R}|$ in the PG topology. Therefore, we drop the terms proportional to the product of $\lambda_{\mathrm{L}}$ and $\lambda_{\mathrm{R}}$ in $D$, and approximate $D$ as
\begin{equation} 
	D \approx  -\frac{(\mathrm{e}^{-g(N-1)}\lambda_{\mathrm{L}} + \mathrm{e}^{g(N-1)}\lambda_{\mathrm{R}} )}{E_{\mathrm{min}}} \sin(\theta)  +\sin(N\theta) \label{D0}.
\end{equation}

To obtain an approximate expression for $E_{\mathrm{min}}$ in terms of $N$,  we assume that $N$ is sufficiently large that one of $\exp(\pm g(N-1))$ is negligibly small. Adopting the approximation that 
\begin{equation}
	\sin(N\theta) \approx \frac{1}{2i} \exp(i N \theta^{(0)} ),
\end{equation} 
(recall that by definition $|\exp(i \theta^{(0)})| \geq |\exp(-i \theta^{(0)})|$) in Eq. \eqref{D0}, approximating $\theta\approx\theta^{(0)}$, and requiring $D=0$ give
\begin{widetext}
    \begin{align}
	& -\lambda_{s_g} \exp(s_g g (N-1)) (2i\sin(\theta^{0})) + E_{\mathrm{min}}\exp(i N \theta^{(0)} ) = 0 \\
	\Rightarrow& \mathrm{ln}\  E_{\mathrm{min}} = N \mathrm{ln}\  ( \exp(s_g g) \exp(- i\theta^{(0)} ) ) + \mathrm{ln}\ ( \lambda_{s_g} \exp(-s_g g) (2i\sin(\theta^{(0)})) ) \label{EMx0}.
\end{align}
\end{widetext}

From the definitions of $g$ and $\theta$ in Eq. \eqref{expg} and \eqref{expitheta}, the terms on the right side of Eq. \eqref{EMx0}  are given by    
\begin{align}
	\exp(s_g g - i \theta^{(0)}) &= \left( -\frac{t_2}{t_1 + s_t s_g \gamma} \right)^{s_t}, \\
	\exp(-s_g g)(2i\sin(\theta^{(0)}) &= -s_t \frac{ t_1^2 - \gamma^2 - t_2^2 }{t_2 (t_1 - s_g \gamma)} \label{e35}.
\end{align} 

Eq. \eqref{linE0} is finally obtained by taking the moduli of the terms on both sides of Eq. \eqref{EMx0}.

Eqs. \eqref{c00} and \eqref{c00a} are obtained by expanding substituting Eqs. \eqref{betaa} and \eqref{betab} in the consistency condition Eq. \eqref{Deq0} and retaining only the lowest power in $E$ the consistency conditions.

\subsection{Appendix B: Bulk states - approaching the PBC spectrum as $N \rightarrow \infty$.}

The numerical results in Fig. \ref{gFig1} suggest that either $|\beta_1|$ or $|\beta_2|$ approaches 1 as $N \rightarrow \infty$ in the bulk states because the eigenenergy spectrum of the bulk states approaches the PBC spectrum, which corresponds to the loci of energy values at which either $|\beta_1|$ or $|\beta_2|$ is equal to 1. We show that this is indeed true. For definiteness, let us consider the case that it is $|\beta_2|$ that approaches 1 so that $|\beta_1| < 1$ in the vicinity of the PBC spectrum. (The argument for the case that $|\beta_1|$ approaches 1 follows analogously.)  In this case, $\beta_1^N \rightarrow 0$ as $N \rightarrow \infty$. In the large $N$ limit, the consistency condition $D=0$ for $D$ given in Eq. \eqref{Dq0} reduces to 
\begin{align}
	& E t_2 \left[E \beta_2^N(t_1+\gamma) - (t_2 + \beta_2(t_1+\gamma) \lambda_{\mathrm{L}} \right]  \nonumber \\
	&+ \frac{1}{\beta_2^2}(t_1+t_2\beta_2-\gamma)\lambda_{\mathrm{L}} \left[ E t_2\beta_2 - \beta_2^N (t_2 + \beta_2(t_1+\gamma)\lambda_{\mathrm{R}}) \right] = 0. \label{bulkX0}
\end{align}
 
Because we assume that $\lambda_{\mathrm{L}}$ and $\lambda_{\mathrm{R}}$ are very small compared to $t_{1,2}$ and that $|\beta_2| \sim 1$, we neglect the term proportional to $\lambda_{\mathrm{L}}$ in the first row of Eq. \eqref{bulkX0} and that proportional to $\lambda_{\mathrm{R}}$ in the second row and solve for $E$ to obtain
\begin{equation}
	E= -\frac{ \beta_2^{1-N} (t_1+t_2\beta_2-\gamma)\lambda_{\mathrm{L}}}{t_1+\gamma} \label{bulkX1}.
\end{equation}
Equating $E^2$ from Eq. \eqref{bulkX1} with the square of the eigenvalue of Eq. \eqref{Hbeta} gives the consistency condition 
\begin{equation}
	\beta_2^{2N+1} = -\left(\frac{\lambda_{\mathrm{L}}}{(t_1+\gamma)}\right)\frac{t_1+t_2\beta_2-\gamma}{ \left(t_2+\beta_2(t_1+\gamma)\right) }. \label{bulkX2}
\end{equation}

We note that except at the two specific values of $\beta_2 = -\frac{t_2}{t_1+\gamma}$ and $\beta_2 = -\frac{\gamma-t_1}{t_2}$, which correspond to the values of $\beta_\pm$ at $E=0$, the right-hand side of Eq. \eqref{bulkX2} has a finite value. This implies that in the  $N\rightarrow \infty$ limit, it is necessary for $|\beta_2| \rightarrow 1$ in order for the left-hand side of Eq. \eqref{bulkX2} to have a finite value so that the consistency condition is satisfied. (Otherwise the left-hand side will shrink to zero [diverge to infinity] in the $N\rightarrow\infty$ limit if $|\beta_2|$ is slightly smaller [larger] than 1.)   This proves that the eigenenergies of the bulk states approach the PBC spectrum, on which every eigenenergy has at least one $\beta$ value with magnitude $|\beta|=1$, as $N \rightarrow \infty$.

\subsection{Appendix C: Derivation of $E_{\mathrm{c}}$ and  $N_{\mathrm{c}}$}

\begin{figure}[htp]
    \centering
    \includegraphics[width=0.48\textwidth]{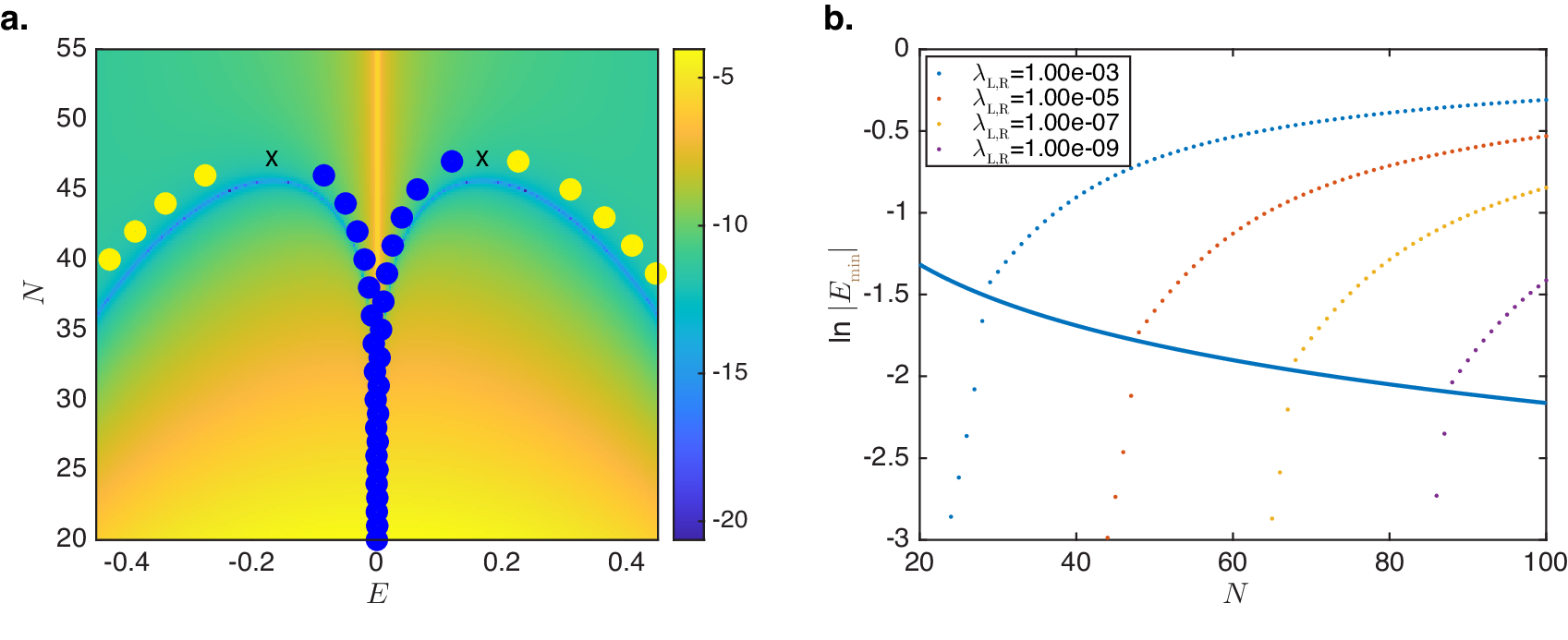}
    \caption{\textbf{a}. $\mathrm{ln}\ \Delta$ plotted as a function of $N$ and $E$ for $t_1=2.5$, $t_2=2.8$, $\gamma=1$, and $\lambda_{\mathrm{L}} = \lambda_{\mathrm{R}} = 10^{-5}$. The circles denote the numerically calculated eigenenergies with the dark blue circles denoting $E_{\mathrm{min}}$ at each given value of $N$ and the green circles the eigenenergies with the second smallest moduli. The X's denote the $(E_{\mathrm{c}},N_{\mathrm{c}})$ points. \textbf{b}.  The loci of $(\mathrm{ln}\ |E_{\mathrm{c}}|$, $N_{\mathrm{c}}$) points calculated using Eq. \eqref{E0x2} and \eqref{M0x1} (continuous lines) for a continuous range of $\lambda_{\mathrm{L}}=\lambda_{\mathrm{R}}$ values and the loci of $(\mathrm{ln}\ |E_{\mathrm{min}}, N)$ points obtained via direct numerical diagonalization for integer $N$ (dots) at $(t_1,t_2)=(2.5,2.8)$, $\gamma=1$ and the $\lambda_{\mathrm{L}}=\lambda_{\mathrm{R}}$ values of $10^{-3}$, $10^{-5}$, $10^{-7}$, and $10^{-9}$. }
    \label{gFig6}
    \end{figure}	
    
We derive $N_{\mathrm{c}}$ and $E_{\mathrm{c}}$ for the $|\beta_1|<|\beta_2|<1$ case in detail, and subsequently describe the extension for the $1 < |\beta_1| < |\beta_2|$ case subsequently. (Recall that this saturation occurs in the PG topology, where both of $|\beta_1|$ and $|\beta_2|$ are either larger than 1 or both smaller than 1.)  $|\beta_1| < |\beta_2| < 1$ implies that  $|\psi^{(\mathrm{A}, \mathrm{B})}(N)| \ll   |\psi^{(\mathrm{A}, \mathrm{B})}(1)|$. We therefore adopt the approximation that $\lambda_{\mathrm{R}}\psi^{(\mathrm{A})}(N) \approx 0$ in Eq. \eqref{bc1} and solve for the coefficient $c$ in $\psi$ the resulting equation. This expression for $c$ is substituted into Eq. \eqref{bc2} and the approximation that $E \psi^{(\mathrm{A})}(N) \approx 0$ adopted to obtain the consistency condition 
\begin{widetext} 
\begin{equation}
	\beta_1 (t_1+\gamma)\lambda_{\mathrm{L}}(\chi_1 - \chi_2) + t_2 \beta_1 \beta_2^{M-2} \left( (t_1+\gamma)\chi_1 - E\right)\chi_2 +  \beta_1^{N-1}t_2\chi_1\left(E-(t_1+\gamma)\chi_2\right) = 0. \label{eqX01} 
\end{equation} 
\end{widetext} 

Dropping the last term proportional to $\beta_1^{N-1}$, which is much smaller than the preceding terms, substituting in Eq. \eqref{chij}, which expresses the $\chi_j$s in terms of the $\beta_j$s, and dropping a term proportional to $E^2$ in the resulting equation, the consistency condition Eq. \eqref{eqX01} approximates to 
\begin{equation}
	\beta_2^{N-2} \approx \frac{ (\beta_2-\beta_1)(t_1+\gamma)\lambda_{\mathrm{L}}}{E(t_1-\gamma)}.  \label{eqX02}
\end{equation}
We denote $|\beta_2^{N-2} - [(\beta_2-\beta_1)(t_1+\gamma)\lambda_{\mathrm{L}}]/(E(t_1-\gamma))|$ as $\Delta$. 

For small and real $|E|$ , the $\beta_1$ and $\beta_2$ values are real, as can be seen from Eq. \eqref{betaa} and \eqref{betab}. $\beta_2$, which appears on the left-hand side of Eq. \eqref{eqX02}, is negative for some parameter ranges. For example, it can be seen from Eq. \eqref{betapm} that $\beta_2$ is negative when $t_2$ and $\gamma$ are positive and $t_1>\gamma$. $\beta_2^{N-2}$ is therefore, in general, complex for real values of $N$ unless $N$ is an integer because a non-integer power of a negative number is complex. Because the right hand side of Eq. \eqref{eqX02} contains only real terms,  Eq. \eqref{eqX02} can only be satisfied at integer values of $N$ where the terms on both sides of the equation are real. Since the sign of $\beta_2^{N-2}$ alternates as $N$ switches between even and odd integer values, $E$ alternates between positive and even values at successive values of $N$, as shown in Fig. \ref{gFig3}a. In contrast, when $\beta_2$ is positive, continuous solutions for $E$ and $N$ exist because a positive number raised to a fractional power is still a positive number. To facilitate the subsequent analysis, we take the modulus of both sides of Eq. \eqref{eqX02}. To achieve reasonably accurate predictions for $E_{\mathrm{c}}$ and $N_{\mathrm{c}}$, it is necessary to expand $\mathrm{ln}\ \beta_2$  to second order in $E$ ($\mathrm{ln}\ \beta_2$ has no linear term in $E$)  and then take the exponent of the expansion, i.e., 
\begin{equation}
	\beta_2^{N-2} \approx \exp( (b_0 + b_2 E^2) \mu)
\end{equation}
where $\mu = (N-2)$ and the specific forms of $b_0$ and $b_2$ are the coefficients in the Taylor expansion of $\mathrm{ln}\ \beta_2$. In contrast, it suffices to retain only the terms proportional to $1/E$ on the right hand side of Eq. \eqref{eqX02}.  Eq. \eqref{eqX02} is then reduced to the general form of 
\begin{equation}
	\exp((b_0 + b_2 E^2) \mu) - \frac{c_0}{E} = 0 \label{eqX03}
\end{equation}
where 
\begin{equation}
	c_0 = \left| \frac{t_1^2-t_2^2-\gamma^2}{t_2(t_1-\gamma)} \right| 
\end{equation}
is the coefficient of the dominant $1/E$ term in Taylor expansion of the right-hand side of Eq. \eqref{eqX02}. 
Fig. \ref{gFig6}a shows that the loci of $\Delta=0$ (the dark blue curve where $\mathrm{ln}\ \Delta$ approaches $-\infty$) calculated under these approximations for an exemplary set of parameters in the PG topology provides a reasonably good approximation to the actual eigenvalues obtained by direct numerical denationalization of the Hamiltonian Eq. \eqref{H0}. In particular, Fig. \ref{gFig6}a also shows that the critical point $(E_{\mathrm{c}}, N_{\mathrm{c}})$ corresponds to a turning point at which $\mathrm{d}\mu/\mathrm{d}E = 0$. To find this turning point, we perform an implicit differentiation of Eq. \eqref{eqX03} with respect to $E$ and obtain 
\begin{equation}
\frac{\mathrm{d}\mu}{\mathrm{d}E} = \left(-\frac{1}{E} - 2 b_2\mu E\right)\left(b_0 + b_2 E^2\right)^{-1}
\end{equation}
from which we read off that at $\frac{\mathrm{d}\mu}{\mathrm{d}E}= 0$,
\begin{equation}
	\mu_{\mathrm{c}} = -\frac{1}{2b_2 E_{\mathrm{c}}^2} \label{mu0} 
\end{equation}
where we now replace $\mu$ and $E$ by $\mu_{\mathrm{c}}$ and $E_{\mathrm{c}}$ in Eq. \eqref{mu0} because whilst Eq. \eqref{eqX01} to Eq. \eqref{eqX03} hold generically for small $|E|$ (e.g., one can see that the \textit{second} smallest real eigenvalues corresponding to each value of $N$ also lie on the $\Delta=0$ curve in Fig. \ref{gFig6}a ),  Eq. \eqref{mu0} holds only at $E = E_{\mathrm{c}}$.  

Substituting the expression for $\mu_{\mathrm{c}}$ in Eq. \eqref{mu0} into Eq. \eqref{eqX03} gives 
\begin{equation}
	\exp\left(-\frac{1}{2}- \frac{b_0}{2b_2 E_{\mathrm{c}}^2}\right) = \frac{c}{E_{\mathrm{c}}}, 
\end{equation}
which yields the solution
\begin{equation}
	E_{\mathrm{c}} = \sqrt{ \frac{b_0}{b_2 \mathcal{W}\left(\frac{ b_0}{b_2 c_0^2 \mathrm{e}}\right) } }. \label{E0x1} 
\end{equation}

Putting in the explicit forms of $b_0$ and $b_2$ gives the general expression for the critical eigenvalue  $E_{\mathrm{c}}$ in Eq. \eqref{E0x2}. The value of $E_{\mathrm{c}}$ obtained using Eq. \eqref{E0x2} can then be substituted into Eq. \eqref{mu0} to obtain $N_{\mathrm{c}} = \mu_{\mathrm{c}} + 2$ as stated in Eq. \eqref{M0x1}.

Having described the derivation of $N_{\mathrm{c}}$ and $E_{\mathrm{c}}$ for the $|\beta_1| < |\beta_2| < 1$ scenario in some detail, we now briefly describe the corresponding derivation for the $1 < |\beta_1| < |\beta_2|$ scenario.  (The $t_1^2 = \pm t_2^2+\gamma^2$ curves form the dividing lines between the two scenarios, as shown in Fig. \ref{gFig2}a. )   In this scenario, $|\psi^{(\mathrm{A}, \mathrm{B})}(M)| \gg  |\psi^{(\mathrm{A}, \mathrm{B})}(1)|$, which leads to the approximations $E \psi^{(A)}(1) \approx 0$ in Eq. \eqref{bc1} and $\lambda_{\mathrm{L}}\psi^{(A)}(1) \approx 0$ in Eq. \eqref{bc2}. Adopting this of approximations, which is analogous to those described above for Eq. \eqref{eqX01}, we obtain the corresponding consistency condition 
\begin{equation}
	\beta_1^{1-N} = \frac{t_2(\beta_2-\beta_1)\lambda_{\mathrm{R}}}{E(t_1-\gamma)}, \label{eqX04} 
\end{equation}
which is the $1 < |\beta_1| < |\beta_2|$ counterpart to Eq. \eqref{eqX02}. 

Comparing Eq. \eqref{eqX04} to Eq. \eqref{eqX02}, we see that the roles of $\beta_2^{N-2}$, $(t_1-\gamma)$ and $\lambda_{\mathrm{R}}$ in  Eq. \eqref{eqX02} in the $|\beta_1| < |\beta_2| < 1$ scenario have been replaced by $\beta_1^{1-N}$, $t_2$, and $t_2$ for the $1 < |\beta_1| < |\beta_2|$  scenario. This correspondence reflects the differences between localization site of the system (i.e., $n=1$ in the former and $n=N+1$ in the latter) and the couplings of this site to its two neighbors along the localization direction.  The corresponding general expressions for $N_{\mathrm{c}}$ and $E_{\mathrm{c}}$ in the $|\beta| > 1$ scenario are hence given by Eq. \eqref{E0x3} and \eqref{M0x3}.

Fig. \ref{gFig6}b shows a comparison between the $(|E_{\mathrm{c}}|, N_{\mathrm{c}})$ values calculated analytically using Eqs. \eqref{E0x2} and \eqref{M0x1} (given by the continuous curve) as $\lambda_{\mathrm{L,R}}$ is varied continuously and the numerically calculated values of $\mathrm{ln}\ |E_{\mathrm{min}}|$ for a few discrete values of $\lambda_{\mathrm{L,R}}$ (represented by the discrete plot points in Fig. \ref{gFig6}b ). This comparison shows that the analytic expressions provide good predictions of $(|E_{\mathrm{c}}|, N_{\mathrm{c}})$ and that the approximations employed are valid. The locus  of the $(\mathrm{ln}\ |E_{\mathrm{c}}|, N_{\mathrm{c}})$ points form a roughly straight line with a negative slope. This indicates that $|E_{\mathrm{c}}|$ varies in an  approximately linear manner with $\lambda_{\mathrm{L},\mathrm{R}}$, which is also hinted at by Eq. \eqref{eqX02} if the variation of $\beta$ with $E$ is neglected.

\subsection*{Acknowledgments}
This paper is supported by the Ministry of Education (MOE) of Singapore Tier-II Grant No. MOE-T2EP50121-0014 (NUS Grant No. A-8000086-01-00) and a MOE Tier-I
FRC grant (NUS Grant No. A-8000195-01-00).


%

\end{document}